\documentclass[aps,prb,twocolumn,floats,showpacs]{revtex4}
\usepackage{epsfig}
\usepackage{color}
\usepackage{bm}
\usepackage{latexsym}

\begin{document}
\newcommand{\hide}[1]{}
\newcommand{\tbox}[1]{\mbox{\tiny #1}}
\newcommand{\half}{\mbox{\small $\frac{1}{2}$}}
\newcommand{\sinc}{\mbox{sinc}}
\newcommand{\const}{\mbox{const}}
\newcommand{\trc}{\mbox{trace}}
\newcommand{\intt}{\int\!\!\!\!\int }
\newcommand{\ointt}{\int\!\!\!\!\int\!\!\!\!\!\circ\ }
\newcommand{\eexp}{\mbox{e}^}
\newcommand{\bra}{\left\langle}
\newcommand{\ket}{\right\rangle}
\newcommand{\EPS} {\mbox{\LARGE $\epsilon$}}
\newcommand{\ar}{\mathsf r}
\newcommand{\im}{\mbox{Im}}
\newcommand{\re}{\mbox{Re}}
\newcommand{\bmsf}[1]{\bm{\mathsf{#1}}}
\newcommand{\mpg}[2][1.0\hsize]{\begin{minipage}[b]{#1}{#2}\end{minipage}}

\title{Transport through quasi-one-dimensional wires with
correlated disorder}
\author{I. F. Herrera-Gonz\'alez, J. A. M\'endez-Berm\'udez, F. M. Izrailev}
\affiliation{Instituto de F\'{\i}sica, Benem\'erita Universidad Aut\'onoma de Puebla,
Apartado Postal J-48, Puebla 72570, Mexico}

\date{\today}

\begin{abstract}
We study transport properties of bulk-disordered quasi-one-dimensional (Q1D) wires paying 
main attention to the role of long-range correlations embedded into the disorder. First, 
we show that for stratified disorder for which the disorder is the same for all individual 
chains forming the Q1D wire, the transport properties can be analytically described provided 
the disorder is weak. When the disorder in every chain is not the same, however, has the 
same binary correlator, the general theory is absent. Thus, we consider the case when only 
one channel is open and all others are closed. For this situation we suggest a 
semi-analytical approach which is quite effective for the description of the total 
transmission coefficient. Our numerical data confirm the validity of our approach. Such Q1D 
disordered structures with anomalous transport properties can be the subject of an 
experimental study.
\end{abstract}

\pacs{03.65.Nk, 	
      73.23.-b	 	
}

\maketitle


\section{Introduction}
Recently, much attention has been paid to the problem of correlated disorder in one-dimensional 
(1D) systems (see for example, Ref.~[\onlinecite{IKM12}] and references therein). As is known, 
in 1D models with white-noise disorder all eigenstates are localized in infinitely large samples, 
independently on the strength of disorder \cite{A58,ATAF80,LR85}. Although this result, known as 
the Anderson localization \cite{A10}, was rigorously proved for continuous potentials long ago, 
one has to note that for the tight-binding and Kronig-Penney models the presence of discrete 
resonances do not allow to treat the problem of localization rigorously for any value of energy 
inside the energy bands. This fact leads to the failure of the single-parameter scaling 
(SPS)\cite{ATAF80} according to which all transport properties of finite samples depend on the 
ratio of the actual sample size to the localization length defined in the limit of infinitely 
large samples. On the other hand, the energy intervals where the SPS fails are typically small 
and in real applications may be neglected.

As recently argued in Refs.~[\onlinecite{SIZC11,HIM13}], outside the resonances the analytical 
results for the transmission coefficient and its variance, obtained for 1D continuous random 
potentials, can be safely used for tight-binding and Kronig-Penny models as well, provided the 
energy is not too close to the resonances. Moreover, it was shown that with specific methods 
one can effectively modify the theory and describe the global properties of the transmission 
within the whole energy region, including the resonance regions \cite{HIM13}.

For a long time the localization properties of weakly random potentials in the presence of 
correlations (colored-noise potentials) were discussed scarcely, in spite of the fact that 
the results developed for continuous 1D potentials have been obtained in the general form 
\cite{LGP88}, allowing to analyze the influence of both short and long-range correlations 
(see discussion in Ref.~[\onlinecite{IKM12}]). The interest in the problem of localization 
for colored-noise potentials has been triggered by numerical studies of discrete dimer models 
for which the onset of delocalization has been observed numerically \cite{F89,DWP90}. Later 
it was understood that such delocalization emerging for discrete values of energies does 
not contradict the general statement of the theory of localization according to which  
delocalization is not possible for 1D disordered potentials. Nevertheless, since in reality 
the size of disordered samples is always finite, one can speak about an {\it effective} 
delocalization within some intervals of energy. Then, the coexistence of localized and 
delocalized eigenstates can be controlled by the specific choice of correlations. This 
non-trivial possibility to control localization properties due to specific correlations has 
led to intensive studies of anomalous transport properties in colored-noise potentials.

One of the results of studies of tight-binding models with colored-noise disorder is the discovery
that long-range correlations can lead to vanishing Lyapunov exponent in a finite range of energy
inside allowed energy bands \cite{IK99,KI99}. Although this result is obtained for weak disorder in 
the first-order perturbation theory, one can speak about the emergence of effective mobility edges 
dividing the regions with localized and extended states. The theoretical predictions of arranging 
narrow energy windows with perfect transmission in a desirable range of energy has 
been experimentally confirmed in microwave experiments with point-like scatterers inserted into 
one-channel waveguides \cite{KIKS00,KIKSU02}. Alternatively, it was shown that with long-range 
correlations one can strongly enhance the localization even when the strength of disorder is weak 
\cite{KIK08}. The important point is that such localization of eigenstates can be achieved in 
quite narrow energy regions, thus resulting in a strong selective reflection of scattering waves. 
Both numerical and experimental results have demonstrated robust anomalous properties of scattering 
even in disordered samples of relatively small size.

It should be stressed that apart from specific colored-noise potentials that can be experimentally
arranged, there are many physical situations where long-range correlations are not avoidable and 
have to be taken into account. One of such situations occurs in experiments of 1D optical lattices 
with interacting bosons (see for example [\onlinecite{So07,Co08,Ro08}]).

In contrast to the problem of correlated disorder in 1D systems for which the theory is practically developed, the transport properties of Q1D systems with colored noise are not well understood. Among the setups studied in this direction one can mention the tight-binding Anderson
model with two-coupled chains \cite{BK07}, the models with stratified or {\it layered}
disorder \cite{IM05,IM04}, and multi-mode waveguides with long-range correlations in surface
profiles \cite{IM03,IM05}.

The aim of this presentation is to contribute to the theory of correlated disorder in the Q1D geometry. Specifically, we consider the Q1D model of the Anderson type in correspondence with the results obtained in the theory of 1D disordered models. As the first step we analyze the situation for which the scattering potential has specific long-range correlations in the longitudinal $x-$direction, however, does not depend on the transverse $y-$coordinate. As is shown in Ref.~[\onlinecite{IM05}], for such {\it stratified} disorder the problem can be solved analytically by the reduction of the Q1D scattering problem to the analysis of the scattering along $N$ independent 1D channels. Specifically, the independence of the disorder on the transverse coordinate allows one to reduce the model to a coset of non-interacting channels characterized by different localization lengths \cite{IM03,TI06}.  Note that since now the total conductance is a complicated combination of partial 1D conductances, the concept of the single parameter
scaling is not valid \cite{IM05,IM04}.

In our study we explicitly show, both analytically and numerically, that long-range correlations
in Q1D wires with stratified disorder can produce a non-monotonic step-wise energy-dependent
conductance. Numerical data indicate that by introducing an additional white-noise disorder in
the transverse direction the effect of long-range longitudinal correlations is strongly suppressed, 
however, it practically remains unaffected in the channel with the lowest index.

This paper is organized as follows. In the next Section we define the model of Q1D wire and give
basic relations for the characterization of scattering properties. Specifically, we show how the
transmission though Q1D wires with stratified disorder can be explained in terms of the independent 
partial transmission coefficients corresponding to the 1D wires composing the Q1D structure. 
In Section III we explain the approach according to which we numerically compute the scattering 
properties of the Q1D correlated wires, based on an effective non-Hermitian Hamiltonian describing 
the scattering. Then, we verify our analytical predictions by comparing them with numerical data, 
and show that for a special choice of the correlated disorder the conductance shows a highly 
unexpected step-wise energy-dependence. Finally, in Section VI we draw our conclusions.

\section{Model and scattering setup}

\begin{figure}[t]
\includegraphics[width=8cm]{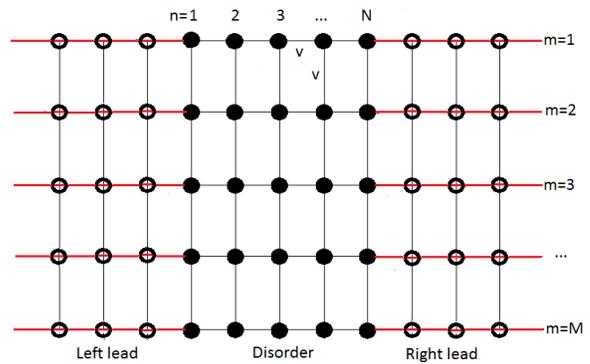}
\caption{(Color online) Disordered Q1D wire of length $N$ and width $M$
connected at both ends to semi-infinite ideal leads.}
\label{Fig1}
\end{figure}

The model consists of a rectangular array of sites of length $N$ and width $M\ll N$, with nearest
neighbor couplings, see Fig.~\ref{Fig1}. The Hamiltonian corresponding to this setup has the
following form,
\begin{eqnarray}
\label{Ham}
\left<n,m\left|H \right|n',m'\right> =
   \epsilon_{nm}\delta_{nn'}\delta_{mm'} - v (\delta_{nn'}\delta_{m,m'+1} && \nonumber \\
+ \ \delta_{nn'}\delta_{m,m'-1} + \delta_{n,n'+1}\delta_{mm'}+\delta_{n,n'-1}\delta_{mm'} ) \ . &&
\end{eqnarray}
The on-site entries $\epsilon_{nm}$ are assumed to be random numbers whose statistical properties
will be specified below, while the coupling amplitudes $v$ between sites are considered to be
constant. The disordered wire is connected to semi-infinite tight-binding leads of width $M$
marked in Fig.~\ref{Fig1} by open circles. In the leads, for $n \leq 0$ and $n \geq N$, the disorder
is absent and the coupling amplitudes between the sites in the leads are also fixed to $v$.

The corresponding stationary Schr\"odinger equation for the eigenstates $\psi_{nm}$ of energy $E$
reads,
\begin{equation}
v (\psi_{n,m+1} + \psi_{n,m-1} + \psi_{n+1,m} + \psi_{n-1,m}) =\left(E-\epsilon_{nm}\right)\psi_{nm} \ .
\label{sto5}
\end{equation}
Without disorder, $\epsilon_{nm}=0$, the solution $\psi_{nm}$ are plane waves with wave
number $\mu_q$ in the transverse direction. In the following, the index $q=1,\dots,M$ is treated as
the channel (or mode) number. In our model we assume zero boundary conditions in the transverse
direction, $\psi_{n,0}=\psi_{n,M+1}=0$, therefore, the dispersion relation takes the form,
\begin{equation}
2v \cos \mu_q=E-2v\cos \left(\frac{\pi q}{M+1} \right) \ , \quad q=1,\dots,M \ .
\label{dis}
\end{equation}

From this relation one can see that the $q-$th channel is open as
along as the energy fulfills the condition inside the interval
\begin{eqnarray}
\label{nm}
-2v\le E-2v\cos \left(\frac{\pi q}{M+1}\right) \le2v \ ,
\end{eqnarray}
and outside it $\mu_q$ becomes imaginary. In fact, the latter equation determines the number $M_1(E)$
of open modes in dependence of the energy. For example, at the band center, $E=0$, all modes are
open and $M_1=M$. While for $|E|<1+\cos\left(\pi/M+1\right)$ all modes are closed. Taking into
account that
\begin{eqnarray*}
\cos \left(\frac{\pi q_1}{M+1} \right)> \cos \left(\frac{\pi q_2}{M+1} \right)
\end{eqnarray*}
for $q_2 > q_1$, one can obtain the expression for the number of open modes in terms of the energy
and channel number $q$,
\begin{eqnarray}
     M_1=
         \left\{
                \begin{array}{lcc}
                  q   &\mbox{if}&  \cos\left(\frac{(q+1)\pi}{M+1}\right)<\frac{|E|}{2v}-1<\cos\left(\frac{q\pi}{M+1}\right) \\
                   0    &\mbox{if}&  \frac{|E|}{2v}>1+\cos\left(\frac{\pi}{M+1}\right)
                \end{array}
         \right. .
\label{open}
\end{eqnarray}

\section{Non-Hermitian Hamiltonian approach}

In order to analyze transport properties of our model in what follows we use the {\it non-Hermitian
Hamiltonian} approach (see for example [\onlinecite{MW69,SZ92,VZ05,Suren2}]). The key point of this approach
is based on the projection of the total Hermitian Hamiltonian $\mathcal{H}$ (disordered part plus leads, see
Eq.~(\ref{Ham})) onto the basis defined by the Hamiltonian $H^{(int)}$ describing the properties of the
{\it closed} model (only disordered part, see Fig.~\ref{Fig1}). In this way the leads are considered
as a continuum to which the disordered part is coupled according to given boundary conditions. The
knowledge of the effective Hamiltonian $H^{(int)}$ allows one to construct the scattering matrix and,
as a result, all transport properties can be obtained.

For our model the non-Hermitian Hamiltonian expressed in the site basis $|n,m\rangle$ has the following
form,
\begin{eqnarray}
\langle n,m|\mathcal{H}(E)|n',m'\rangle=\langle n,m|H^{(int)}|n',m'\rangle \nonumber \\
-\sum^{M_1}_{q=1}\mbox{e}^{i\mu_q}\left( \gamma_L\delta_{n1}+\gamma_R\delta_{nN}\right)\delta_{nn'}P_{qm'}P_{qm} \ .
\label{po}
\end{eqnarray}
Here $H^{(int)}$ is the Hermitian Hamiltonian of the internal system, while the second term in the
right hand side corresponds to the coupling of the internal system to the leads. In the general case
the coupling is characterized by two parameters, $\gamma_L$ and $\gamma_R$, with $L$ and $R$ denoting
the left and right leads, respectively. In our study, for simplicity, we assume symmetric couplings,
$\gamma_L=\gamma_R=\gamma$. As is defined above, $\mu_q$ stands for the wave number at the $q-$th
channel and is related to the energy $E$ through the dispersion relation (\ref{dis}). The elements
$P_{ij}$ in Eq.~(\ref{po}) are nothing but the basis states in which the Hamiltonian $H^{(int)}$ is
presented. In fact, they are eigenstates of this Hamiltonian in the absence of disorder,
\begin{eqnarray}
P_{ij} & = & \sqrt{\frac{2}{M+1}}\sin \left(\frac{\pi i j}{M+1} \right).
\label{pl}
\end{eqnarray}

Equation (\ref{po}) can be written in matrix form as follows,
\begin{equation}
{\bf \mathcal{H}}(E)={\bf H^{(int)}}+2\pi {\bf A}{\bf Q}(E){\bf A}^T-i\pi {\bf A}{\bf A}^T \ .
\label{NH}
\end{equation}
Here ${\bf H^{(int)}}$ is the $NM \times NM$ Hermitian matrix with ordered matrix elements
$\Big \langle n,m\left|H^{int}(E)\right|n',m'\Big \rangle$. The second Hermitian and third
non-Hermitian terms in the right hand side of Eq.~(\ref{NH}) represent the real and imaginary
parts of the coupling to the leads, respectively. As for the coupling matrix ${\bf A}$ of
size $MN \times 2M_1$, it is composed by the ordered coupling amplitudes
${\bf A}=\{A^{(Lq)}_{mn},A^{(Rq)}_{mn} \}$ between the internal states $|n,m\rangle$ and open
left and right channels $Lq$ and $Rq$, respectively. Thus, the {\it coupling amplitudes} are
given by
\begin{eqnarray}
\label{amp}
A^{Lq}_{mn} & = & (\gamma_L/\pi)^{1/2}\sqrt{\sin \mu_q}P_{qm}\delta_{n1} \ , \\ \nonumber
A^{Rq}_{mn} & = & (\gamma_R/\pi)^{1/2}\sqrt{\sin \mu_q}P_{qm}\delta_{nN} \ .
\end{eqnarray}
${\bf Q}(E)$ is the $2M_1 \times 2M_1$ diagonal matrix with real elements ordered as
$\{Q_1,Q_2,\dots,Q_{M_1},Q_1,Q_2,\dots,Q_{M_1}\}$, where
\begin{equation}
Q_q(E)=-\left(\cot \mu_q \right)/2, \ \ q=1,\dots, M_1(E).
\end{equation}

Now, from the effective non-Hermitian Hamiltonian we can pass to the scattering ${\bf S}$-matrix
written in the channel space,
\begin{equation}
{\bf S}= \left(\begin{array}{cc}
      {\bf t} & {\bf r'} \\
      {\bf r} & {\bf t'}
    \end{array} \right) =
\frac{1+{\bf {\cal C}}^{\dag}{\bf{\cal K}}}{1+{\bf{\cal C}{\cal K}}} \ ,
\label{sm1}
\end{equation}
where ${\bf t}$, ${\bf t'}$, ${\bf r}$, and ${\bf r'}$ are $M_1\times M_1$ transmission and
reflection matrices. Below, we chose the coupling parameter $\gamma$ in such a way that the
transmission in each channel is maximal. This means that we consider the so-called {\it perfect
coupling} for which the average scattering matrix is zero $\langle S\rangle=0$. In our case,
both for the 1D model with $M=1$ and for the Q1D model with $M>1$, the perfect coupling corresponds
to $\gamma\approx v$.\cite{Suren2}

It can be shown that the matrix ${\bf {\cal C}}$ in Eq.~(\ref{sm1}) of size $2M_1\times 2M_1$
has the following structure,
\begin{eqnarray*}
{\bf{\cal C}}=i\pi-2\pi {\bf Q}(E).
\end{eqnarray*}
As for the {\it reaction matrix} ${\bf{\cal K}}$ (of the same size $2M_1\times 2M_1$), its matrix
elements are defined by
\begin{eqnarray}
\label{ka4}
{\cal K}_{ab}(E) & = & \sum_{nm}\frac{\widehat{A}^{(a)}_{nm}\widehat{A}^{(b)}_{nm}}{E-E_{nm}} \ ,  \\
\widehat{A}^{(a,b)}_{nm} & = & \sum_{rs} A^{(a,b)}_{rs}\psi_{rs}(E_{nm}) \ . \nonumber
\end{eqnarray}
Here, $\psi_{rs}$ are the components of the eigenvector of the matrix $\bf{H^{int}}$
having the eigenvalue $E_{nm}$, and we have introduced the channel index $a,b\equiv cq$
that indicates which lead $c = L, R$ and mode $q$ we refer to. Once the $S$-matrix is known one
can calculate the dimensionless conductance $g=\left(2e^2/h\right) T$, where
 $T = \mbox{Tr} (tt^\dagger)$ is the transmission coefficient\cite{Landauer}, with $e$ as the
charge of the electron and $h$ as the Planck constant.

It should be noted that at the band center the relation (\ref{NH}) reduces to the simple form,
\begin{eqnarray*}
{\bf \mathcal{H}}={\bf H^{(int)}}-i\pi {\bf A}{\bf A}^T ,
\end{eqnarray*}
in which the coupling to continuum is described by the imaginary term  only. In addition, in the 1D case
the scattering matrix (\ref{sm1}) takes the well known form,
\begin{eqnarray*}
{\bf S}=\frac{1-i\pi{\bf {\cal K}}}{1+i\pi{\bf {\cal K}}} \ .
\end{eqnarray*}

\section{correlated stratified disorder}
\subsection{Analytical results}

In this Section we consider the so-called stratified disorder for which the potential
is independent of the transverse coordinate quantized by the index $m$, i.e.
\begin{eqnarray}
\epsilon_{n1}=\epsilon_{n2}=\dots=\epsilon_{nM}\equiv \epsilon_n.
\label{stra}
\end{eqnarray}
In this case our model can be reduced to a set of $M$ 1D independent chains which are nothing
but 1D tight-binding Anderson models\cite{IM05}. To show this, first, we rewrite the Schr\"odinger
equation (\ref{sto5}) in the matrix form,
\begin{eqnarray}
{\bf a}^{(n-1)}+{\bf a}^{(n+1)}=\left({E}-{\bf B}^{(n)}-{\bf C}\right){\bf a}^{(n)} \ ,
\label{sto}
\end{eqnarray}
where ${\bf a}^{(n)}$ is the vector with components $\psi_{nm}$ ($m=1,\dots,M$).
Here ${\bf B}^{(n)}$ and ${\bf C}$ are $M \times M$ matrices with elements given by
\begin{eqnarray}
B^{(n)}_{ij} & = & \epsilon_{nj}\delta_{ij} \ , \nonumber \\
C_{ij} & = & v(\delta_{i,j+1} + \delta_{i,j-1}) \ .
\end{eqnarray}
Then, we pass to a new unperturbed basis through the transformation,
\begin{eqnarray}
{\bf b}^{(n)}&=&{\bf P}{\bf a}^{(n)} \ ,
\end{eqnarray}
where the columns of the matrix ${\bf P}$ are the eigenvectors of the Hamiltonian matrix
${\bf C}$ that correspond to the 1D Anderson model of size $M$ with vanishing disorder,
$\epsilon_i=0$, and zero boundary conditions. Note that the corresponding eigenvectors
and eigenvalues are analytically known. In this representation the Schr\"odinger equation (\ref{sto})
takes the form,
\begin{eqnarray}
{\bf b}^{(n+1)}+{\bf b}^{(n-1)}=\left(E-{\bf P}^T{\bf B}^{(n)}{\bf P}-{\bf D} \right){\bf b}^{(n)} \ ,
\label{sto2}
\end{eqnarray}
where the elements of the $M \times M$ matrix ${\bf D}$ are given by
\begin{eqnarray}
D_{ij} & = & 2v\cos \left(\frac{\pi i}{M+1} \right)\delta_{ij} ,
\label{PP}
\end{eqnarray}
while the elements of the $M \times M$ matrix $\bf P$ are given by Eq.~(\ref{po}).

For the stratified disorder Eqs.~(\ref{sto2}) become uncoupled since
${\bf P}^T{\bf B}^{(n)}{\bf P}={\bf B}^{(n)}$ is a diagonal matrix. Hence, the Q1D
Anderson model is reduced to a set of $M$ 1D chains, where the energy for each
chain is given by
\begin{eqnarray}
E_q = E-2v\cos \left(\frac{\pi q}{M+1} \right) \ , \quad q=1,\dots,M \ .
\label{shiftE}
\end{eqnarray}

Let us now specify the properties of the stratified disorder. First, we assume
zero mean and small variance $\sigma^2$ of the site energies,
\begin{eqnarray}
\langle\epsilon_n \rangle=0 \ , \quad \sigma^2=\langle\epsilon^2_n \rangle \ll 1 \ ,
\label{weak}
\end{eqnarray}
where $\langle\cdots\rangle$ denotes the average over different realizations
of disorder. Apart from that, we assume that the statistical properties of disorder
are defined by the two-point correlator describing the long-range correlations.
Therefore, an additional ingredient of the disorder is the specific form of the
normalized binary correlator,
\begin{equation}
\chi(k) = \frac{\left<\epsilon_{n} \epsilon_{n+k}\right>}{\sigma^2} ,
\end{equation}
to be defined below.

Since our model with the stratified disorder can be rigorously reduced to a set of $M$
1D chains, one can try to apply the theory of 1D localization developed for continuous
potentials. According to this theory for weak disorder and in the limit
$N \rightarrow \infty$, the eigenstates $b^{(n)}_q$ (in our case, the components of
the vectors ${\bf b}^{(n)}$, see Eq.~(\ref{sto2})) are exponentially localized with
the characteristic length $l^{(q)}_{\infty}$ related to the $q-$th channel. As is
known, the inverse localization length can be defined in terms of the Lyapunov exponent
$\lambda_q$, and the analytical expression for the latter has the following form
(see for example Ref.~[\onlinecite{IKM12}]),
\begin{eqnarray}
\label{lia3}
\lambda_q &\equiv& \frac{1}{l^{(q)}_{\infty}} =  \frac{\sigma^2}{8\sin^2 \mu_q}W(2\mu_q) \ , \nonumber \\
W(2\mu_q) & = & 1+2\sum^{\infty}_{k=1} \chi(k) \cos{2\mu_q k} \ .
\end{eqnarray}
As one can see, the correlation properties of the random sequence $\{\epsilon_n\}$ are
entirely defined by the power spectrum $W(\mu)$ of the binary correlator. Note that this
result is correct for weak disorder; as for the higher moments of the correlations they
may contribute to the localization length only in the next order of perturbation theory
in the disorder parameter $\sigma^2$. Note that for white noise disorder, we have $W(2\mu_q)=1$.

Now we focus on the problem of scattering through the disordered region represented by full
circles in Fig.~\ref{Fig1}. Since for the stratified disorder the transmission along every 1D
channel is independent from those along the other channels, the total transmission coefficient $T$ can be expressed as the sum of
partial coefficients $T_q$ corresponding to the propagation of incident plane waves along the
$q-$th independent channels,
\begin{equation}
T(E)=\sum^{M_1}_{q=1}T_q(E) \ .
\label{sum}
\end{equation}
This expression agrees with the Landauer concept of conductance \cite{Landauer}.
Here $M_1(E)$ is the number of open modes in the leads given by Eq.~(\ref{open}).

Above we announced that we plan to use the analytical results developed for 1D {\it continuous}
models. However, the model under consideration
is a discrete model for which the theoretical analysis is restricted due to the presence of
resonances in the energy space (see for example [\onlinecite{IKM12,THI12}] and references therein).
Specifically, there are no rigorous results for the probability distribution of $T$, or
equivalently, for the corresponding moments for any energy inside the allowed energy band. To the contrary,
for continuous random potentials with weak disorder the problem of scattering through finite
1D wires was rigorously solved by various analytical approaches \cite{IKM12,LGP88}. In
particular, there is an exact expression for the average transmission coefficient in terms on
the ratio of the localization length $l_{\infty}$ to the length $N$ of the sample \cite{IKM12,LGP88},
\begin{eqnarray}
\langle T_q\rangle&=&\sqrt{\frac{2x^3_q}{\pi}}\mbox{exp}\left(-\frac{1}{2x_q} \right)\int^{\infty}_0\frac{z^2}{\cosh z}\mbox{exp}\left(-\frac{z^2x_q}{2} \right)dz , \nonumber \\
x_q &=& l^{(q)}_{\infty}/N \ .
\label{Ta}
\end{eqnarray}
In Eq.~(\ref{Ta}), we have added the index $q$ in order to indicate to which channel we are
referring to. Here, the brackets stand for the average over a number of different realizations
of the correlated disorder.

As was found in Ref.~[\onlinecite{SIZC11}], the latter expression turns out to be very good even for the
1D tight-binding Anderson model, provided the energy values are not very close to the resonances.
Moreover, recently a new approach has been developed in Ref.~[\onlinecite{HIM13}] allowing one to modify
the standard perturbation theory in such a way that it gives good results also at the resonant
energies. In particular, it was shown that Eq.~(\ref{Ta}) still gives a good description of
numerical results at the resonances, provided the expression for the localization length takes
into account the influence of these resonances. Thus, relation (\ref{Ta}), together with
Eqs.~(\ref{lia3}) and (\ref{sum}), give us the possibility to obtain the expression for the
transmission coefficient $\langle T(E)\rangle$ in dependence on the model parameters. Below we
verify the validity of Eq.~(\ref{Ta}) by comparing it with numerical data.

\subsection{Numerical data}

Since the localization length $l^{(q)}_{\infty}$ of any $q-$th conducting channel is fully
determined by Eq.~(\ref{lia3}), if the power spectrum $W(2\mu_q)$
vanishes within some energy window, the corresponding channel will be fully transparent in
this energy interval \cite{IK99,KI99,IM09}. This prediction has been confirmed both numerically
and experimentally for the 1D Anderson model (see details and references in Ref.~[\onlinecite{IKM12}]).
Here our interest is on how this effect can be seen in the Q1D model with the stratified disorder.
To do this, we choose the following form of the binary correlator,
\begin{eqnarray}
\chi(k)=\frac{1}{2k(\mu_R-\mu_L)}\left(\sin 2\mu_Rk-\sin 2\mu_Lk\right) \ .
\label{corr}
\end{eqnarray}
As one can see, this correlator exhibits a power law decay which is typical for long-range
correlated disorder. It can be shown that the correlator (\ref{corr}) results in the step-wise
power spectrum,
\begin{eqnarray}
    W(E_q)=
         \left\{
                \begin{array}{ll}
                  W_0 & \ \mbox{if} \ E_L\le \left|E_q\right|\le E_R \\
                  0 & \ \mbox{if} \ |E_q|<E_L \ \mbox{or}\ E_R<|E_q|\le 2v
                \end{array}
         \right. .
\label{nosen}
\end{eqnarray}
Here, $\mu_L$ and $\mu_R$ are related to $E_L$ and $E_R$ through the dispersion law for the 1D system,
$E=2\cos \mu$, and $W_0$ is determined by the normalization condition $\int^{\pi/2}_0 W(\mu)d\mu=\pi/2$
(see details in Ref.~[\onlinecite{IKM12}]). Note that we have omitted the transverse index $m$ since each
chain has {\it the same} disorder sequence $\epsilon_n$. In our numerical simulations we consider
the following fixed values: $E_L=0.4$, $E_R=1.3$, $\sigma^2=0.02$, and $v=1$.

\begin{figure}[t]
\includegraphics[width=7.0cm]{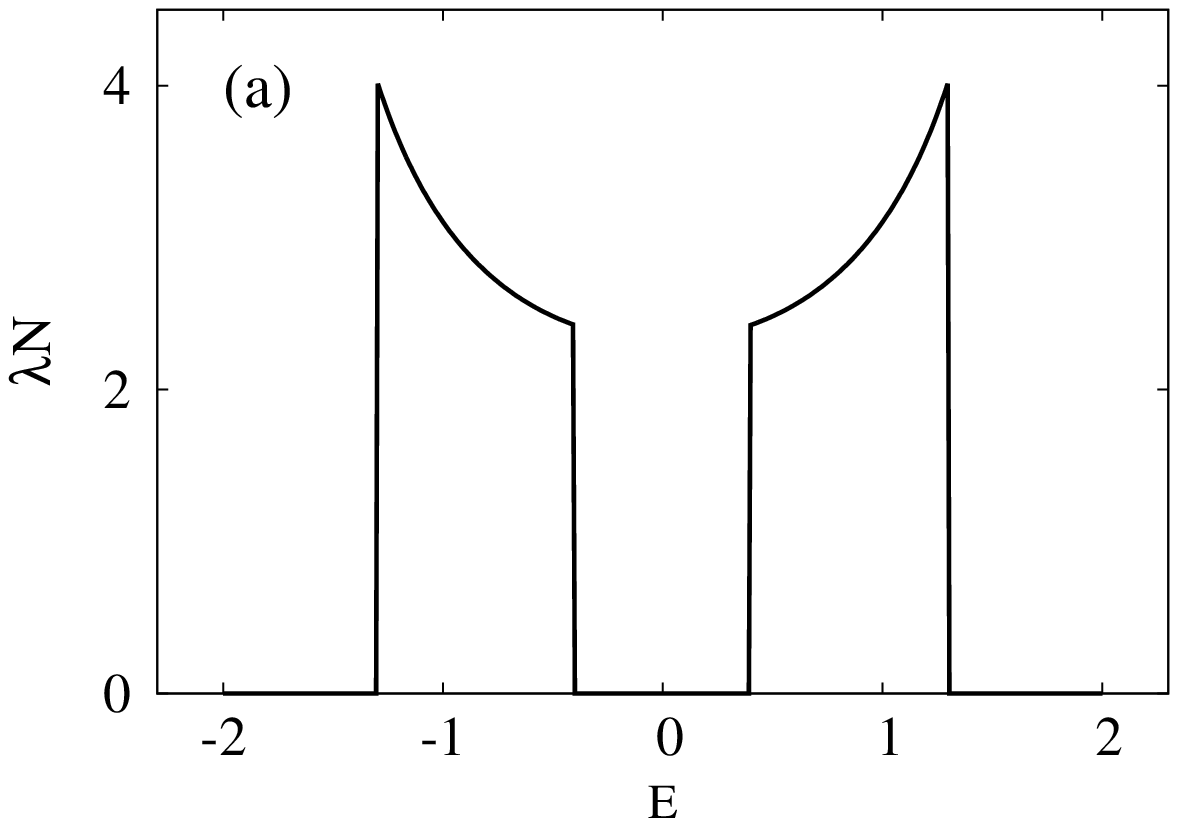}
\includegraphics[width=7.0cm]{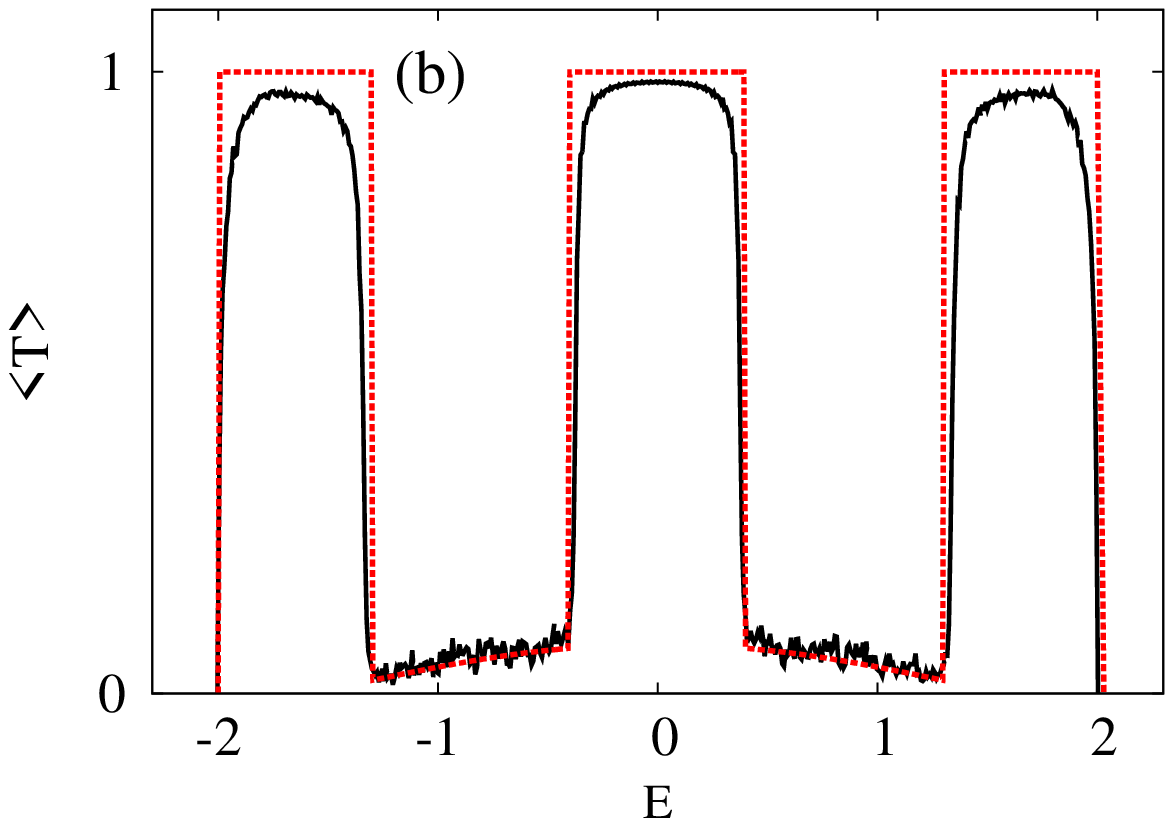}
\caption{(Color online) (a) Rescaled Lyapunov exponent $\lambda N$ as a function of the energy $E$ for one chain,
$M=1$. The correlated disorder has been fixed by the step-wise power spectrum of Eq.~(\ref{nosen}) with
$E_L=0.4$, $E_R=1.3$, $\sigma^2=0.02$, and $v=1$. (b) Average transmission coefficient $\bra T \ket$
as a function of $E$. The continuous curve corresponds to numerical data while the dashed one is the
theoretical prediction from Eqs.~(\ref{lia3}-\ref{Ta}). The average is taken over 100 realizations of
disorder for a disordered region of length $N=300$.}
\label{Fig2}
\end{figure}

Correlations with the power spectrum (\ref{nosen}) have been already used in the literature to control
the transport properties of 1D systems \cite{IM09,ivan2}. In Fig.~\ref{Fig2} we demonstrate these properties
by considering our model with one chain only, $M=1$. Specifically, we present the analytical expression
for the Lyapunov exponent (\ref{lia3}), together with the predicted and actual dependence of the
transmission coefficient $\langle T \rangle $, see Eq.~(\ref{Ta}). One can see a good correspondence
with the data showing the expected windows of transparency in dependence of the energy $E$.

\begin{figure}[t!]
\includegraphics[width=7cm]{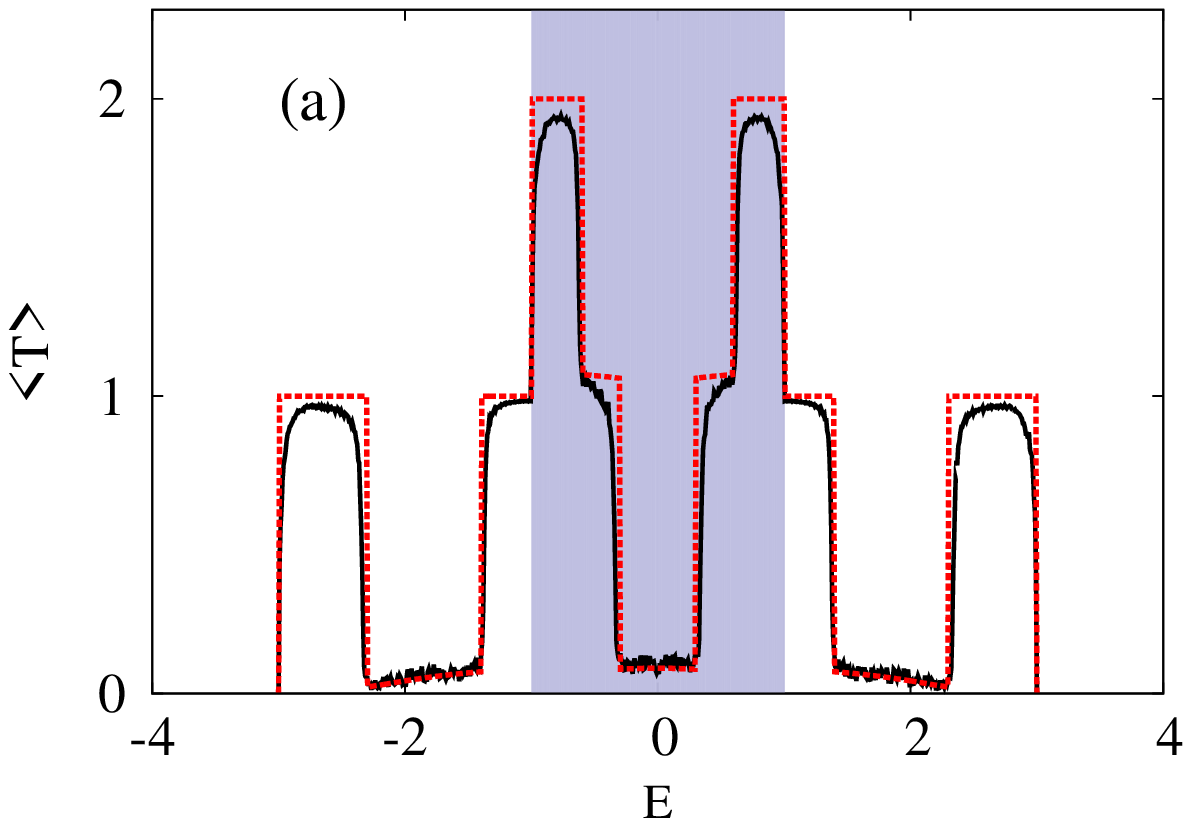}
\includegraphics[width=7cm]{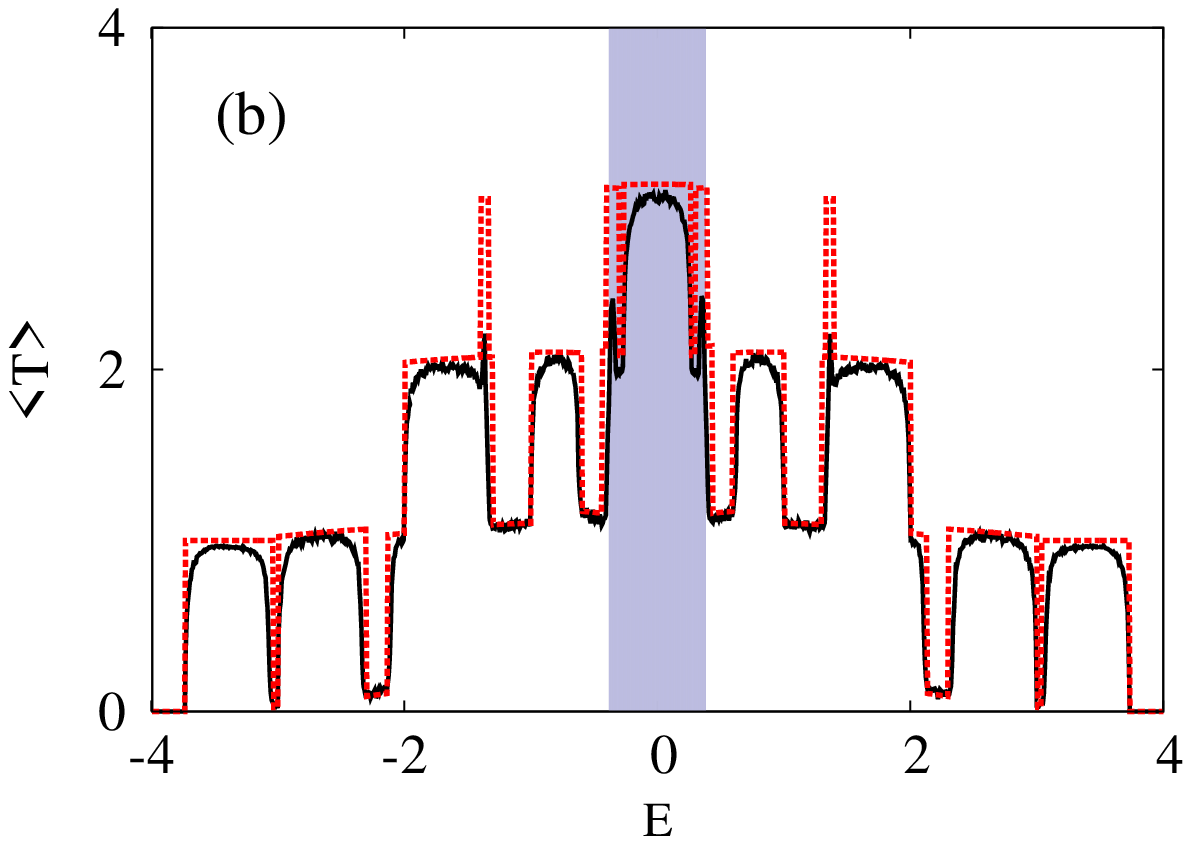}
\includegraphics[width=7cm]{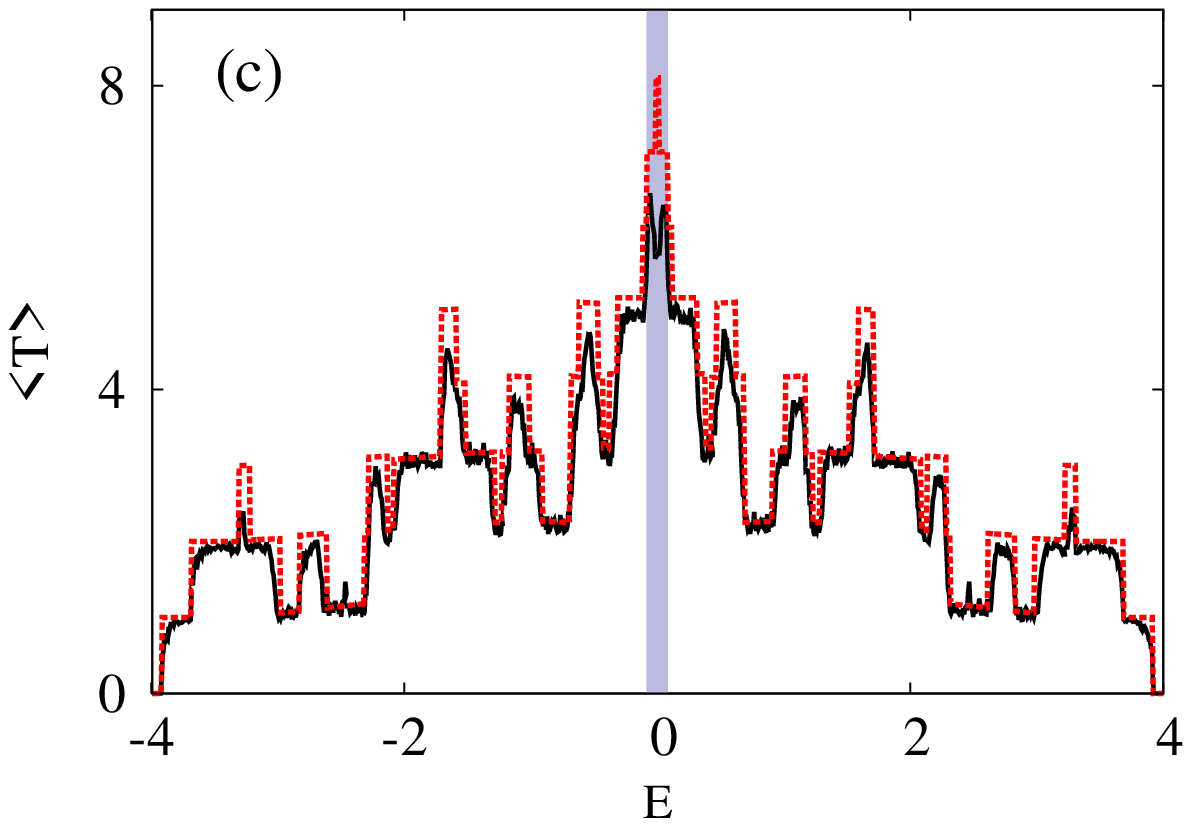}
\caption{(Color online) Same as in Fig.~\ref{Fig2}(b) for the Q1D model with 
(a) $M=2$, (b) $M=5$, and (c) $M=10$. The energy window where all modes in the 
leads are open is shown by the shaded region.}
\label{Fig3}
\end{figure}

For the Q1D model with the stratified disorder the scenario is much more complicated. Indeed, for any of
the $q-$th channels the windows of transparency are defined by the energy shifted in accordance with
Eq.~(\ref{shiftE}). Specifically, for each channel there are three transparent energy windows given
by Eq.~(\ref{nosen}). Outside of these windows the transmission vanishes due to the chosen type of
correlations. Thus, the total transmission is obtained by the overlap of the energy dependencies
corresponding to each channel. One can show that the average transmission coefficient is approximately
defined by the integer number due to the following expression,
\begin{eqnarray}
N_c(E)&=&\sum^M_{q=1}\Big ( \Theta[E_q+E_L]-\Theta[E_q-E_L]+\Theta[E_q-E_R]\nonumber \\
&-&\Theta[E_q+E_R]+\Theta[E_q+2v]-\Theta[E_q-2v] \Big) ,
\label{nc}
\end{eqnarray}
with $\Theta[x]$ as the Heaviside step function. The resulting step-wise behavior is shown in
Figs.~\ref{Fig3}. Notice that the maximum value of the average
conductance is reached, in most of the cases, in the energy region where without disorder all modes
are open, therefore, the average transmission takes its maximum value $M$. The general properties of
such a behavior were predicted in Ref.~[\onlinecite{IM05}]. As one can see, the main feature of the
correlated disorder is the non-monotonic dependence of the transmission in dependence on the energy.

\begin{figure}[t!]
\includegraphics[width=7cm]{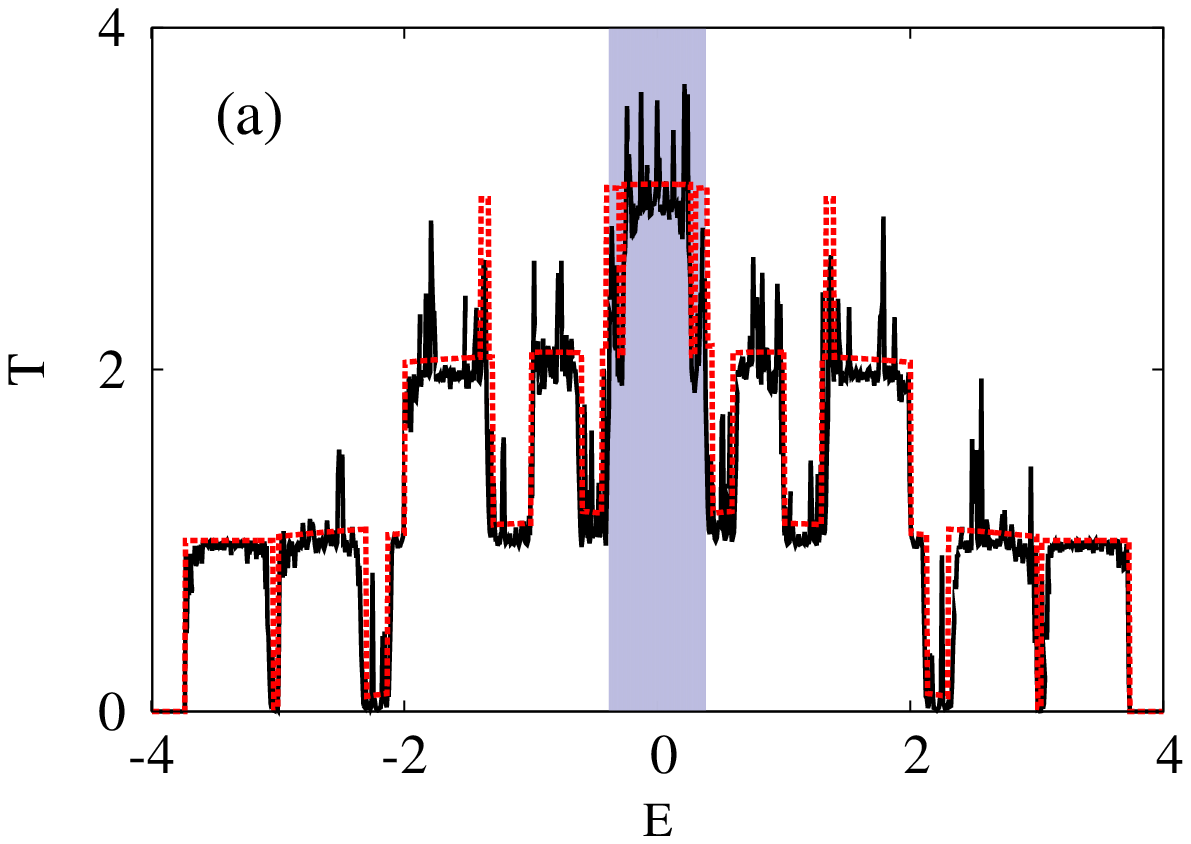}
\includegraphics[width=7cm]{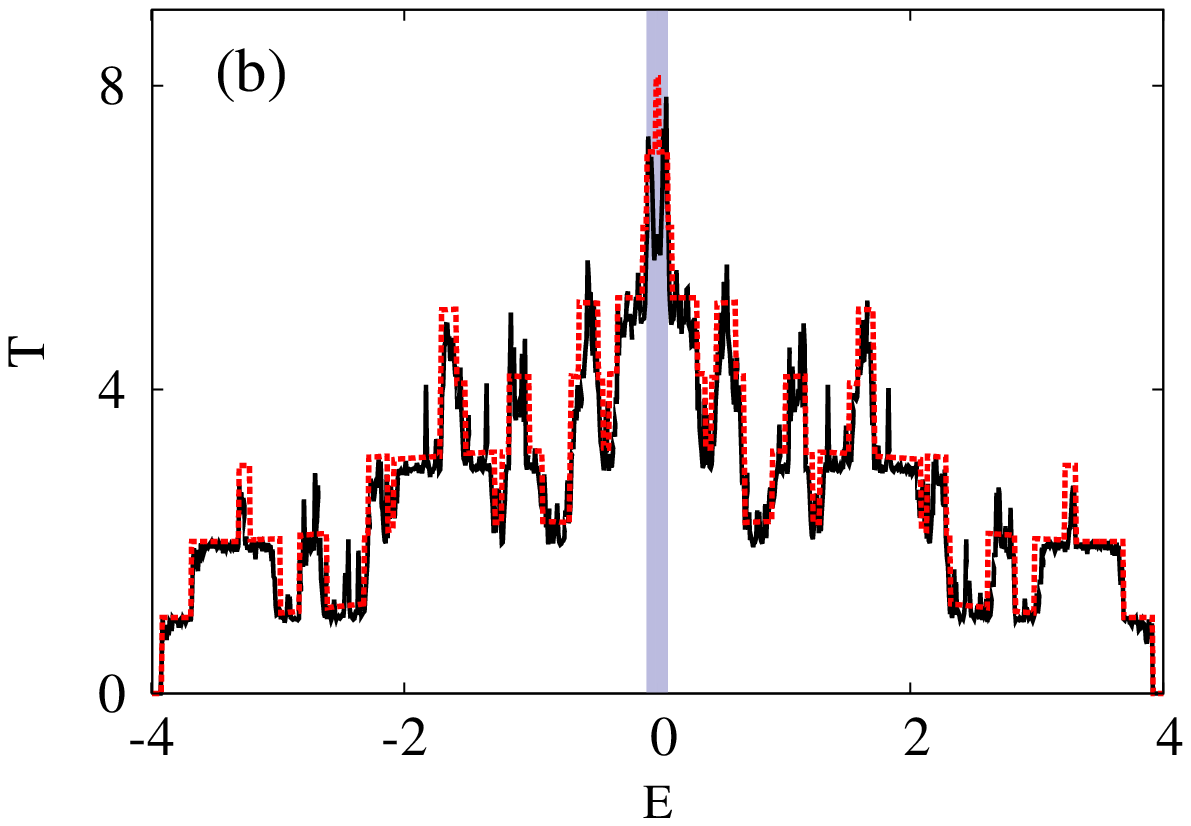}
\caption{(Color online) Transmission coefficient $T$ as a function of $E$ for the Q1D model of length
$N=300$ with (a) $M=5$ and (b) $M=10$. Continuous curves correspond to the numerical data while dashed
curves are the theoretical predictions from Eqs.~(\ref{lia3}-\ref{Ta}). A single realization of disorder
was used. The energy window where all modes in the leads are open is indicated by the shaded region.}
\label{Fig5}
\end{figure}

The data presented in Figs.~\ref{Fig3} is obtained by averaging over a large number
of disorder realizations with the same kind of long-range correlations. However, from the experimental
view point it is important to know whether the non-monotonic dependence of $T(E)$ can be practically
seen for individual samples. To answer this question, it may be helpful to understand how the fluctuations of the transmission
coefficient for an individual disordered sample of size $N$ depend on the model parameters. According
to the theory of 1D disordered systems in the ballistic regime, $N/l^{(q)}_{\infty}\ll 1$, the variance
of $T_{q}$ can be written as follows \cite{IKM12}, 
\begin{equation}
\mbox{Var}(T_{i})=4\left(\frac{N}{L^{(i)}_{\infty}}\right)^2+
O\left(\left(\frac{N}{L^{(i)}_{\infty}}\right)^3\right) \ .
\end{equation}
Therefore, one can expect that if the $q-$th channel is open (ballistic regime, $T_q \approx 1$), the
fluctuations should be not very strong. On the other hand, if due to specific long-range correlations
the $q-$th channel is closed (localized regime, $T_q \ll 1$) the fluctuations are also not important.
The numerical data reported in Fig.~\ref{Fig5} demonstrate that the non-monotonic step-wise dependence of
$T(E)$ can be still detected. However, the fluctuations can wash out very narrow peaks; this can be clearly seen
when the average is performed (compare with Fig.~\ref{Fig3}).

\section{Correlated non-stratified disorder}

Another important problem refers to the Q1D model (\ref{Ham}) with the disorder having the same
correlation properties in the longitudinal direction for each chain, however, with no correlations
in the vertical direction. Mathematically, this means that the disorder potential depends on both
transverse and longitudinal coordinates, being correlated along the sample, however, completely
uncorrelated transverse to the sample. In other words, the statistical properties for the site
energies are defined as follows,
\begin{eqnarray}
\langle \epsilon_{nm}\rangle=0 \ ,  \ \ \sigma^2=\langle \epsilon^2_{nm}\rangle \ , \ \
\chi(k) \delta_{mm'}=\frac{\langle \epsilon_{nm}\epsilon_{n+k,m'}\rangle}{\sigma^2} \ , \nonumber \\
\label{sta}
\end{eqnarray}
where $\chi(k)$ is the normalized binary correlator of the site energies. As before, we assume
that the disorder is weak, $\sigma^2 \ll 1$. In the numerical calculations the specific form of
the correlator is chosen according to Eq.~(\ref{corr}) with the corresponding power spectrum
(\ref{nosen}). However, the analytical approach can be applied to systems with any form of the
binary correlator $\chi(k)$.

When the correlated disorder is not stratified, the model can not be represented by a set of independent 1D
systems, and the analytical solution for the localization length is known for specific cases only.
It should be, however, noted that in the case of weak white-noise disorder the problem has been
solved in Refs.~[\onlinecite{VG,VG1}]. As for the correlated disorder, the general solution has been
obtained for the two-chain model only \cite{BK07}.

Our particular interest in this Section is in the situation when one channel is open and all the
others are closed. Only in this case we can suggest the phenomenological expression for the
Lyapunov exponent, and make the comparison with numerical data. Thus, below we consider the
Q1D model under the following conditions corresponding to this situation,
\begin{eqnarray}
2v\left(1+\cos \frac{2\pi}{M+1}\right)<\left| E\right|<2v\left(1+\cos \frac{\pi}{M+1}\right),
\label{op}
\end{eqnarray}
where the first mode $q=1$ is open if $E>0$. Note that if $E<0$ the open mode is the last one,
with $q=M$. For the two-chain model with one open channel the formula for the inverse localization
length reads \cite{BK07}, 
\begin{eqnarray}
\lambda(E)&=&\frac{1}{32\sin^2 \mu_q}\left[\left<\epsilon^2_{n1}\right> W_{11}(2\mu_q)
+\left<\epsilon^2_{n2}\right> W_{22}(2\mu_q) \right.  \nonumber \\
&+&\left. 2\left<\epsilon_{n1}\epsilon_{n2}\right>W_{12}(2\mu_q) \right],
\label{two}
\end{eqnarray}
where $\mu_q$ is the wave number of the open mode ($q=1,2$). 
Here the power spectra $W_{ij}(\mu)$ are defined as
\begin{eqnarray*}
W_{ij}(\mu)=1+2\sum^{\infty}_{k=1}\chi_{ij}(k)\cos (\mu k) \ , 
\end{eqnarray*}
with the binary correlators $\chi_{ij}(k)$ given by
\begin{eqnarray*}
\chi_{11}(k)&=&\frac{\langle\epsilon_{1,n}\epsilon_{1,n+k}\rangle}{\langle\epsilon^2_{n1}\rangle} \ , \ \
\chi_{22}(k)= \frac{\langle\epsilon_{2,n}\epsilon_{2,n+k}\rangle}{\langle\epsilon^2_{n2}\rangle} \ ,\nonumber \\
\chi_{12}(k)&=&\frac{\langle\epsilon_{1,n}\epsilon_{2,n+k}\rangle}{\langle\epsilon_{n1}\epsilon_{n2}\rangle} \ .
\end{eqnarray*}
Formula (\ref{two}) is valid for weak disorder and not very close to the critical energies 
where the transition from propagating to evanescent regime occurs.

Let us apply Eq.~(\ref{two}) to our model with two chains having the statistical properties 
given by Eq.~(\ref{sta}). Therefore, $W_{ij}(2\mu_q)=W(2\mu_q)$, and
\begin{equation}
\lambda(E)=\frac{\sigma^2W(2\mu_1)}{16\sin^2\mu_1}.
\label{chan}
\end{equation}
Note that the localization length is a symmetric function with respect to $E=0$. Therefore, 
all localization properties are the same either for $q=1$ or $q=M$.

Another analytical result, however, for the white-noise disorder reads \cite{VG1},
\begin{eqnarray}
\lambda(E)=\frac{(3+\delta_{2,M+1})\sigma^2}{16(M+1)\sin^2\mu_1} ,
\label{chan1}
\end{eqnarray}
which is written for $M$ chains and one open channel. With the use of this expression, one 
can suggest the phenomenological generalization valid for the correlated disorder as well,
\begin{equation}
\lambda(E)=\frac{(3+\delta_{2,M+1})\sigma^2W(2\mu_1)}{16(M+1)\sin^2\mu_1}.
\label{final}
\end{equation}
Note that in the case of uncorrelated disorder, $W(2\mu_q)=1$, the latter expression reduces 
to Eq.~(\ref{chan1}) and for $M=2$ the result (\ref{chan}) is recovered. Quite remarkable is 
the prediction that the inverse localization length decreases as the number of longitudinal 
chains $M$ increases.
\begin{figure}[t]
\includegraphics[width=7.0cm]{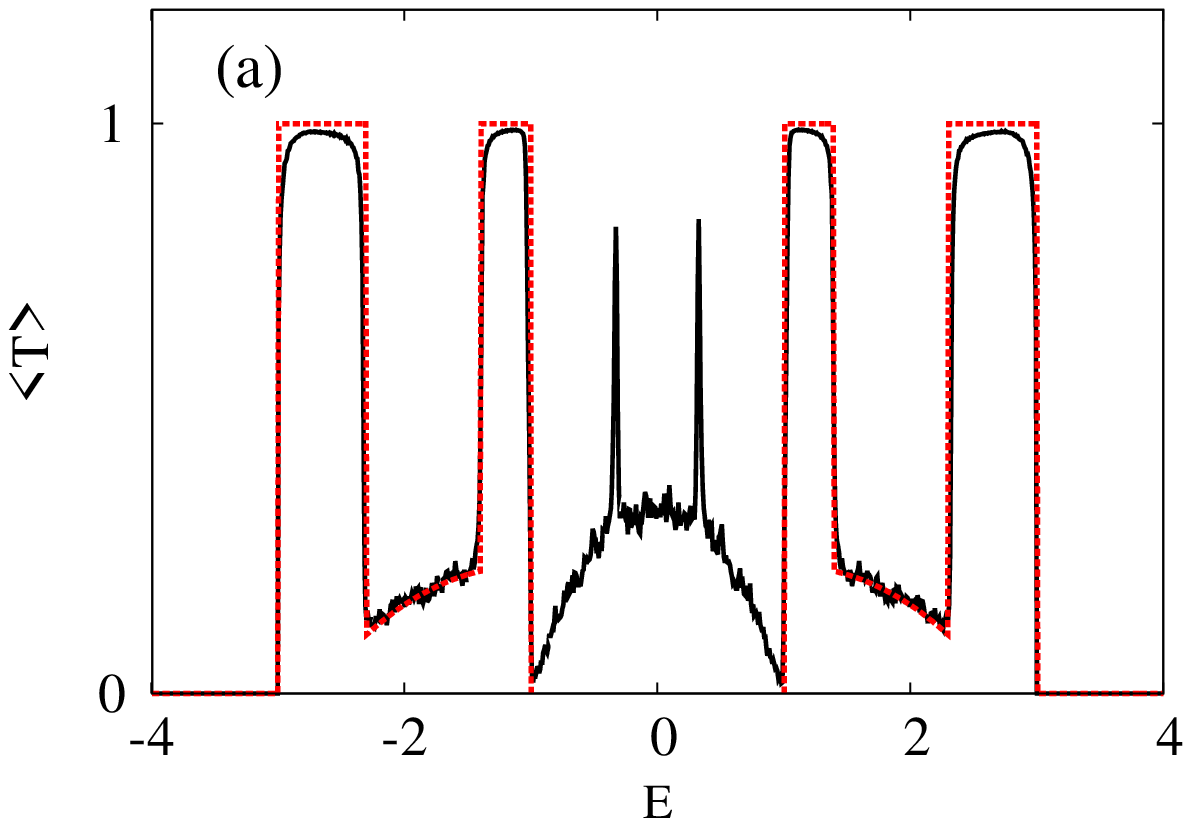}
\includegraphics[width=7.0cm]{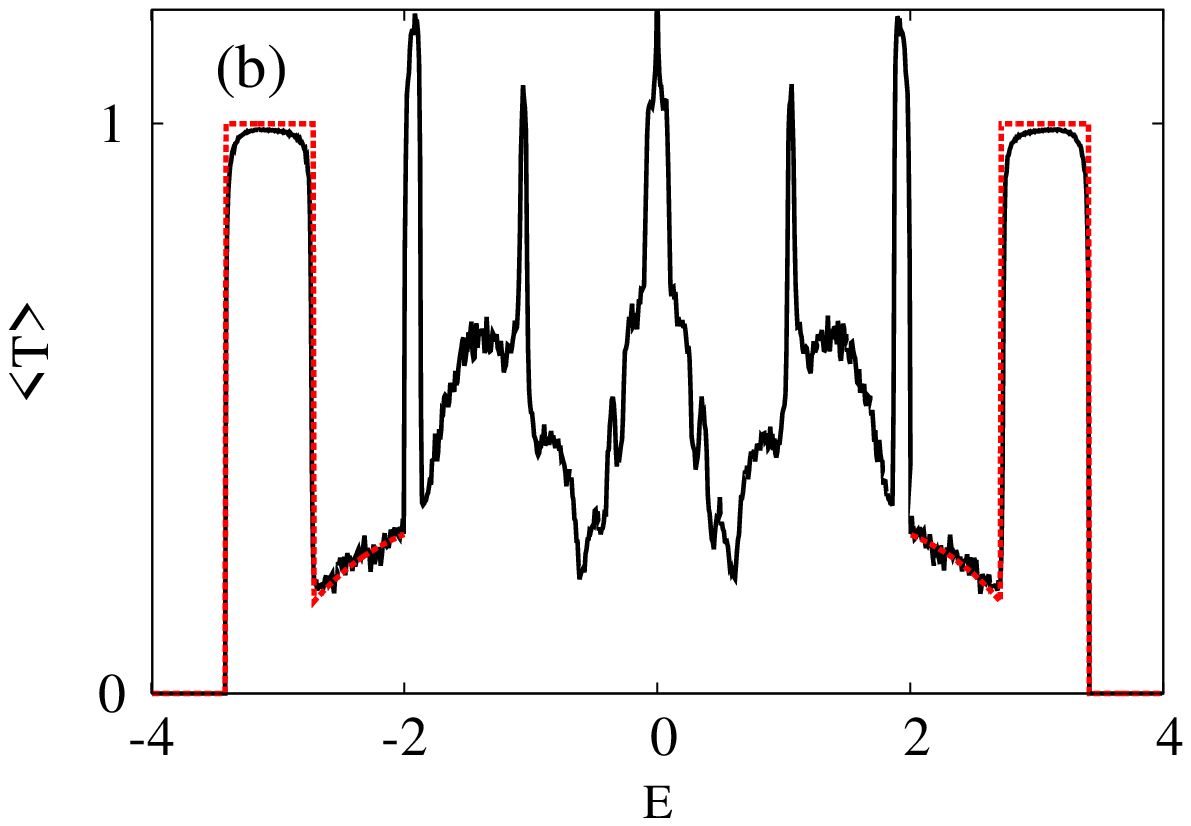}
\includegraphics[width=7.0cm]{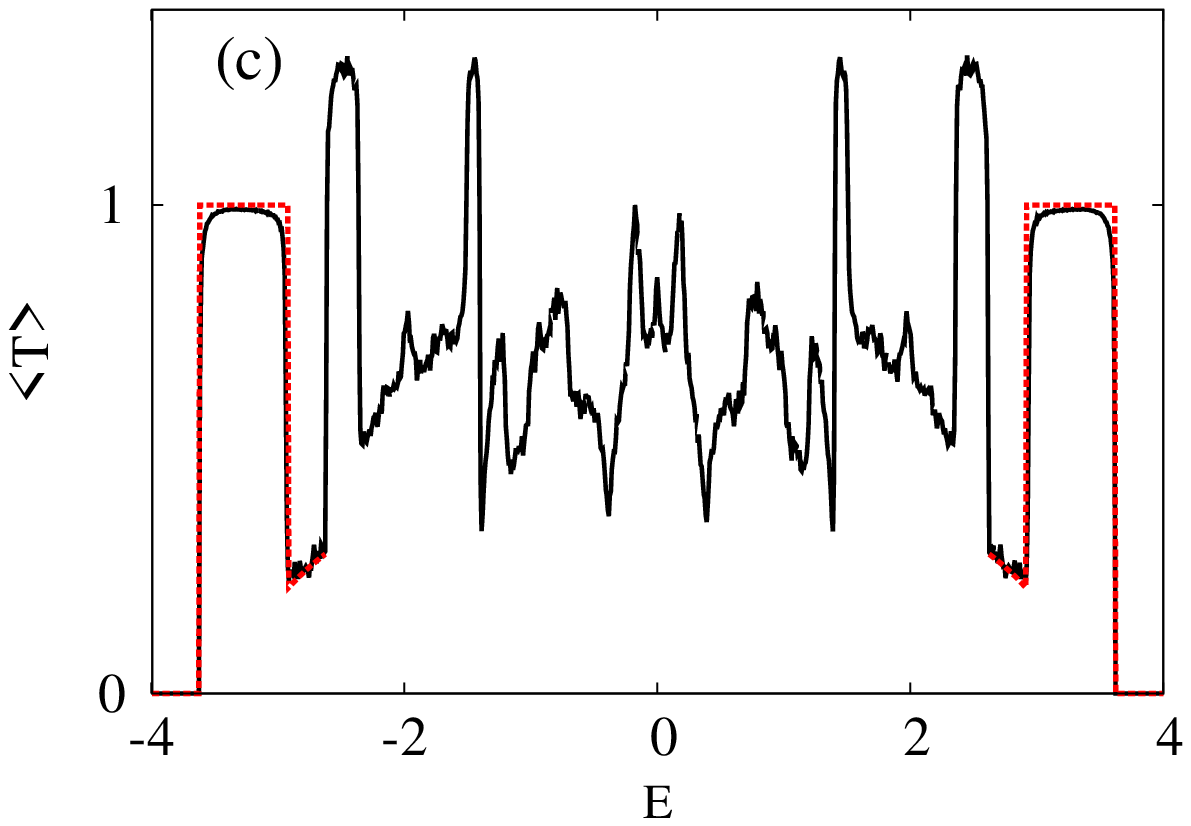}
\caption{\label{figu15} (Color online) Average transmission coefficient as a function of the 
energy for non-stratified disorder. Continuous curves represent the numerical data and dashed
curves are the theoretical prediction given by Eqs.~(\ref{Ta}) and (\ref{final}). The length of
the system is $N=300$ with the disorder strength $\sigma^2=0.02$ for (a) $M=2$, (b) $M=3$, and
(c) $M=4$.}
\end{figure}
It should be pointed out that here we study the case when the longitudinal and transverse
hopping amplitudes are equal, see Fig.~\ref{Fig1}. Therefore, all the sub-bands (\ref{nm}) are overlapped.
However, the expression (\ref{final}) can be generalized even for the case when the transverse hopping
amplitudes are not equal to the longitudinal ones, this could lead to the non-overlapping of some
sub-bands. In this case one has also to take into account the influence of the evanescent modes as
stated in Ref.~[\onlinecite{He}]. Another result can be found in Ref.~[\onlinecite{ZZ}] where the
localization-delocalization transition was studied when considering the strength of the transverse
hopping as an independent parameter.

Now, since only one open mode contributes to transmission, one can suggest that the expression for
the average total transmission (in the energy region where only one mode is open) can be obtained
by inserting the localization length defined by Eq.~({\ref{final}}) into Eq.~(\ref{Ta}). Indeed, our
numerical data presented in Figs.~\ref{figu15} manifest that Eq.~({\ref{final}}) gives very good
description for the total transmission coefficient in the energy regions corresponding to only one 
open channel. One can also see that when more than one channel is open the energy dependence of the 
transmission coefficient acquires a quite complicated form, thus indicating the failure of the 
analytical approach in the general case.

\section{Conclusions}

We have studied the transport properties of bulk-disordered Q1D wires paying attention to the role
of long-range correlations along disordered structures of finite size. First, we have manifested
that in the case of stratified disorder all transport properties can be fully explained analytically.
As predicted in Ref.~[\onlinecite{IM05}], in this case the expression for the total transmission coefficient
$T$ can be presented as a sum of partial coefficients $T_q$ that correspond to independent 1D chains
characterized by the index $q$. Since the theory of correlated disorder for 1D tight-binding models
is fully developed (see for example [\onlinecite{IKM12}]), this allows one to incorporate the obtained
results into the problem of the correlated transport for Q1D disordered systems. Our numerical data
demonstrate a perfect agreement with the analytical predictions. For the numerical study we have used
the approach which is based on the non-Hermitian Hamiltonians from which one can construct the
scattering matrix, therefore, all transport characteristics can be extracted.

As the second step, we have analyzed the model in which the long-range correlations are taken in the
same way as for the stratified disorder, however, the individual realizations of the disorder in the
chains are independent from each other. Since in this case the disorder potential depends on both
coordinates, the longitudinal and transverse ones, there is mixing between different channels when
the waves propagate through the Q1D structure. Since the general theory is absent, we have studied
the situation when one channel is open only, while $M-1$ other channels are closed. For this case the
rigorous theory is also absent, however, we suggest an approach which gives a phenomenological
expression for the transmission coefficient, based on the results obtained earlier for white-noise
disorder. The suggested expression turns out to be very good, as the comparison with the data shows.
It should be noted that such a situation when the transport in Q1D systems is practically defined by
one open channel can be easily arranged experimentally.

Finally, we would like to point out that specific long-range correlations can result in a strong
enhancement of the localization, even when the disorder is weak (see results, discussion, and
references in Ref.~[\onlinecite{IKM12}]). This effect is clearly seen from our numerical data that is
obtained for relatively short disordered samples. Specifically, the emergence of the energy windows
where the transmission coefficient vanishes or becomes very small, is a direct consequence of the
enhancement of the localization. Interestingly enough, such an enhancement of localization in the
selected energy windows is accompanied by the suppression of localization in the complementary
energy windows within the energy band\cite{IKM12}. Therefore, the long-range correlations can be
considered as the mechanism for the redistribution of the degree of localization in the energy space.
This effect can be used to manufacture devices with controlled transport properties in photonic
heterostructures, semiconductor superlattices, and electron nanoconductors; among others.

\begin{acknowledgments}
F.M.I. acknowledges the support from CONACyT Grant No.~N-133375 as well as from the VIEP-BUAP
grant IZF-EXC13-G. J.A.M.-B. acknowledges support from VIEP-BUAP grant MEBJ-EXC14-I and from
PIFCA BUAP-CA-169.
\end{acknowledgments}

\end{document}